\documentclass[twocolumn,showpacs,preprintnumbers,amsmath,amssymb]{revtex4}
\usepackage{graphicx}
\usepackage{dcolumn}
\usepackage{bm}

\begin{document}

\title{On $V_{ud}$ determination from kaon decays}

\author{Mihail V. Chizhov}
\affiliation{Centre for Space Research and Technologies, Faculty of Physics,\\
University of Sofia, 1164 Sofia, Bulgaria
}%


\begin{abstract}
The pion $\beta$ decay $\pi^+\to\pi^0 e^+\nu$ proceeds through
pure weak vector hadronic currents and, therefore, the theoretical
prediction for it is more reliable than for the processes with
axial-vector current contribution. For example, recently the pion
$\beta$ decay has been used for $V_{ud}$ determination. The main
aim of this letter is to point that kaon $\beta$ decay $K^0\to
K^+(\pi^+\pi^0)e^-\bar{\nu}$ analogously can be used for this
purpose.
\end{abstract}


\maketitle

\section{Introduction}

Flavor transitions within and between different quark generations
due to the weak charged interactions are described by the
Cabibbo--Kobayashi--Maskawa (CKM) unitary
mixing-matrix~\cite{Cabibbo,KM}.
\begin{equation}\label{CKM}
 V_{CKM}=\left(
 \begin{array}{ccc}
   V_{ud} & V_{us} & V_{ub} \\
   V_{cd} & V_{cs} & V_{cb} \\
   V_{td} & V_{ts} & V_{tb} \
 \end{array}
 \right)
\end{equation}
This approach is founded on a pure phenomenological basis and the
determination of the matrix elements is completely based on
experimental data. If different types of experiments provide a
consistent values for a particular matrix element, this should point
to the correctness of our results. Therefore, new ideas about
systematically independent measurements are welcome.

Often all new things are well forgotten old ones, however, from
time to time they appear in the scientific literature without
citing the sources. The subject that we will present concerns the
determination of the well defined matrix element $V_{ud}$.
Therefore, it could be sometimes and somewhere discussed. We
apologize for eventual plagiarism, nevertheless, we would like to
point to a new possibility to measure this quantity in kaon
$\beta$ decays.

Up to now the most precision determination of this value follows
from the superallowed $0^+\to 0^+$ nuclear $\beta$-decay
experiments~\cite{Hardy,Savard}. Based on the data from over 100
different experiments and using a new method for controlling
hadronic uncertainties in the radiative correction to superallowed
nuclear beta decays along with refinements from \cite{CMS},
Marciano and Sirlin~\cite{Vud} have derived the adopted at present
PDG value~\cite{PDG}
\begin{equation}\label{Vud}
  \vert V_{ud}\vert = 0.97377\pm0.00027.
\end{equation}

However, a recent modern determination of the
$Q$-value~\cite{Savard} of the superallowed decay of the
radioactive nuclei $^{46}$V, obtained from the mass difference of
$^{46}$V and its decay daughter $^{46}$Ti, gives a new $Q$-value
and invalidates the set of its seven previous measurements. This
value affects the evaluation of $V_{ud}$ from superallowed nuclear
decays and leads to a somewhat lower value for $V_{ud}$. It may
indicate a problem with $Q$-values of the other superallowed
emitters used for $V_{ud}$ determination.

Therefore, independent determination of $V_{ud}$ from other
experiments is needed. The second precise evaluation of $V_{ud}$
value, with bigger than superallowed transition uncertainties, is
obtained from the measurements of the neutron lifetime and the
$\beta$ asymmetry coefficient $A$~\cite{PERKEO}. The later
measurements are necessary in order to fix the unknown
contribution of the axial-vector nuclear matrix element into the
neutron decay rate, which is the main source of uncertainty for
$V_{ud}$ extraction.

Using the most precise updated value for the ratio $\lambda^{\rm
exp}=g_A/g_V=-1.2733(13)$~\cite{Mund} of the axial coupling
constant to the vector coupling constant and the PDG value for the
neutron lifetime $\tau_n=885.7(8)$ s, one can evaluate $V_{ud}$
as~\cite{Vud}
\begin{equation}\label{nVud}
  \vert V_{ud}\vert=0.97218\pm0.00101 .
\end{equation}
This value is 1.5$\sigma$ lower than the extracted one (\ref{Vud})
from superallowed nuclear decays and may indicate the presence of
new interactions. Their effect, predicted in the \cite{neutron},
leads to the corrected value of $\lambda=-1.2714(13)$ and to more
consistent $\vert V_{ud}\vert=0.97339(101)$ value.

However, the extracted from the neutron decays $V_{ud}$ value
depends on the experimentally measured neutron lifetime, for which
present situation is unclear due to the most recent
result~\cite{Serebrov} $\tau_n=878.5(8)$. Nevertheless, precision
and consistent $\lambda$ determinations from several correlation
coefficient measurements, which are ongoing and planned, would
indicate reliable experimental results and would be able to put
more stringent constraints on new physics.

And the last but not least important source of information about
$V_{ud}$ is the very clean theoretically $0^-\to 0^-$ pion $\beta$
decay $\pi^+\to\pi^0 e^+\nu$. It is a pure vector transition and
is free from nuclear structure uncertainties. However, due to the
small pion mass difference it has a very weak branch, of the order
of 10$^{-8}$, which leads to severe experimental difficulties.
Nevertheless, the PIBETA Collaboration~\cite{PIBETA}, using the
Paul Scherrer Institute facilities, has improved the experimental
uncertainty for this mode up to 0.6\% and quotes
\begin{equation}\label{pVud}
  \vert V_{ud}\vert=0.9728\pm0.0030 .
\end{equation}
Therefore, to reach the precision of $V_{ud}$ determination from
superallowed nuclear decay (\ref{Vud}) tenfold improvements both
in statistics and systematics are necessary. One hopes that with
the development of high-intensity proton drivers, this aim can be
reached.

It is worth nothing here, that besides excellent possibility for
pion and muon physics, these facilities give a unique possibility
for kaon physics as well. For example, a CP violation beyond the
Standard Model can be searched in rare kaon decays with branching
ratios $10^{-10}-10^{-12}$. One of the background processes is
$0^-\to 0^-$ kaon $\beta$ decay $K^0\to K^+e^-\bar{\nu}$, which
can give an additional information about $V_{ud}$ value. We are
going to discuss this in the next section.

\section{Kaon beta decays}

The kaon $\beta$ decay $K^0\to K^+e^-\bar{\nu}$ is completely
analogous to the pion beta decay $\pi^-\to\pi^0 e^-\bar{\nu}$. It
can serve as a possibility to extract $V_{ud}$ matrix element,
because the strange quark $s$ does not participate in the weak
interactions and play a spectator role. As far as the final kaon
is not a stable particle, it can be registered through its decay
modes. From our point of view, the most probable decay channel
$K^+\to\mu^+\nu$ is not appropriate for the final state
identification, because the two neutral neutrinos escape from
registration. Therefore, pure hadronic modes, mainly
$K^+\to\pi^+\pi^0$ decays, are very suitable for this.

It is interesting to note, that the experimental signature of
these decays $K^0\to\pi^+\pi^0 e^-\bar{\nu}$ does not fulfill
$\Delta S = \Delta Q$ selection rule, in contrast to the allowed
$K_{e4}$ decays $K^0\to\pi^-\pi^0 e^+\nu$, but indicates the
presence of $\Delta S = -\Delta Q$ weak transitions. This
situation is completely analogous to the experimental puzzle of
the beginning of sixties with the observation of $\Sigma^+\to
n\mu^+\nu$ decays~\cite{1962}, which later have been realized as
background~\cite{Cronin}.

So, let us consider the background events
\begin{equation}\label{Kdecay}
  K^0\to K^+ e^-\bar{\nu}\to\pi^+\pi^0 e^-\bar{\nu}
\end{equation}
to get a valuable information about the first CKM matrix element.
In order to obtain competitive with (\ref{pVud}) result, we need
to keep all uncertainties of the order of $10^{-3}$. The amplitude
of the first semileptonic decay in (\ref{Kdecay})
\begin{equation}\label{MK}
  {\cal M}=-\frac{G_F}{\sqrt{2}} V_{ud}\left[
  f_+(q^2)(p+p')_\mu+f_-(q^2)q_\mu\right]\ell^\mu
\end{equation}
is expressed through the form factors $f_+$ and $f_-$ of hadronic
matrix element $\langle K^+(p')\vert\bar{u}\gamma_\mu d\vert
K^0(p)\rangle$ multiplied by the leptonic current
\begin{equation}\label{l}
  \ell^\mu=\bar{e}\gamma^\mu(1-\gamma^5)\nu.
\end{equation}

In general, the form factors depend on the square of the momentum
transfer to the lepton pair $q_\mu=(p-p')_\mu$. However, even for
the pion $\beta$ decay the Dalitz plot integral is practically
insensitive to this dependence~\cite{Cirigliano}, and the form
factors can be considered as constants. We can also neglect the
form factor $f_-$, which is proportional to the small isospin mass
difference $m_{\pi^+}^2-m_{\pi^0}^2$, and in $SU(2)$ symmetry
limit is equal to zero. Moreover, it is multiplied on the momentum
transfer $q_\mu$, which effectively leads to the small
contribution in the Dalitz plot distribution, proportional to
$(m_e/m_{K^0})^2\approx 10^{-6}$.

Furthermore, for kaon $\beta$ decay, in which the initial and
final hadrons belong to an $I=1/2$ multiplet, $f_+=1$ with good
precision. Isospin corrections in first non-zero approximation are
given by the formula~\cite{ChPT}
\begin{equation}\label{df}
  \delta f_+=H_{\pi^+\pi^0}+2H_{K^+K^-}\approx -6.5\times 10^{-6}.
\end{equation}
They are negligibly small in accordance to the Ademollo-Gatto
theorem~\cite{AD}.

Therefore, the rate of kaon $\beta$ decay is given by the
well-known formula~\cite{tK}
\begin{equation}\label{tKbeta}
  \frac{1}{\tau_{K\beta}}=\frac{G_F^2}{60\pi^3}\vert
  V_{ud}\vert^2\left(1-\frac{\Delta}{2m_{K^0}}\right)^3\Delta^5
  f(\epsilon,\Delta)(1+\delta),
\end{equation}
where
\begin{equation}\label{DK}
  \Delta=m_{K^0}-m_{K^+}=3.972\pm 0.027~{\rm MeV}
\end{equation}
is kaon mass difference~\cite{PDG}, $\epsilon=(m_e/\Delta)^2$, and the Fermi
function $f$ is given by
\begin{eqnarray}\label{f}
  f(\epsilon,\Delta)=\sqrt{1-\epsilon}&&\hspace{-0.4cm}
  \left[1-\frac{9}{2}\,\epsilon-4\epsilon^2
  +\frac{15}{2}\epsilon^2\ln\left(\frac{1+\sqrt{1-\epsilon}}{\sqrt{\epsilon}}
  \right)\right.\nonumber\\
  &&-\left.\frac{3}{7}\frac{\Delta^2}{(m_{K^0}+m_{K^+})^2}\right],
\end{eqnarray}
while $\delta$ represents the effect of the radiative corrections.

The second decay in (\ref{Kdecay}) is pure hadronic decay with an
experimental branching ratio~\cite{PDG}
\begin{equation}\label{Kp2pi}
  B(K^+\to\pi^+\pi^0)=(20.92\pm0.12)\%.
\end{equation}
Therefore, the rate of $K^0\to\pi^+\pi^0 e^-\bar{\nu}$ decay
(\ref{Kdecay}) can be estimated as
\begin{equation}\label{K02pi}
  \frac{1}{\tau_{K^0\to\pi^+\pi^0 e^-\bar{\nu}}}
  =\frac{B(K^+\to\pi^+\pi^0)}{\tau_{K\beta}}
  \approx 0.02~\frac{1}{\rm s}
\end{equation}
This decay channel has been proposed to measure the kaon mass
difference $\Delta$, due to its clear signature and very high
sensitivity of the kaon $\beta$ decay to the latter~\cite{Jaffe}.

However, $K^0$ and $\bar{K}^0$ states are not invariant under CP
symmetry transformation and do not represent physical states.
Instead $\vert K_S\rangle = p\vert
K^0\rangle+q\vert\bar{K}^0\rangle$ and $\vert K_L\rangle =
p\,\vert K^0\rangle-q\vert\bar{K}^0\rangle$ combinations are
assigned to the physical mesons $K_S$ and $K_L$, respectively. In
the case of CP invariance, $q=p=1/\sqrt{2}$, $K_S$ represents CP
even and $K_L$ CP odd states. Using PDG fit value for mean life of
$K_L$ meson~\cite{PDG}
\begin{equation}\label{tKL}
  \tau_{K_L}=(5.114\pm 0.021)\times 10^{-8}~{\rm s}
\end{equation}
one can estimate the branching ratio of the sum of the two-step
decay (\ref{Kdecay}) and its CP transformed
\begin{equation}\label{KLdecay}
   K_L\to\pi^+\pi^0 e^-\bar{\nu}+\pi^-\pi^0 e^+\nu
\end{equation}
as
\begin{equation}\label{BK}
  B(K_L\to\pi^\pm\pi^0 e^\mp\bar{\nu}(\nu))
  =\frac{\tau_{K_L}}{\tau_{K^0\to\pi^+\pi^0 e^-\bar{\nu}}}
  \approx 10^{-9}.
\end{equation}

\section{Discussion and conclusions}

In this section we would like to discuss the experimental
possibility of $V_{ud}$ extraction from decays (\ref{KLdecay})
with an accuracy not worse than the one from pion $\beta$ decays.
As far as this decays have very small branching ratio, one needs
high-intensity beam provided in average 10$^7$  $K_L$-decays
per second. Probably, the best place for such measurements is the
50 GeV Proton Synchrotron at JHF. In order to derive the rate of
the two-step process (\ref{Kdecay}) from (\ref{BK}) one needs to
know with a good precision the lifetime of $K_L$ meson or some of
its branching ratios. The present accuracy of the lifetime
(\ref{tKL}) $\delta\tau_{K_L}/\tau_{K_L}\approx 4\times 10^{-3}$
is already good enough and contributes to $V_{ud}$ error at the
level of $2\times 10^{-3}$.

The uncertainty in the determination of the rate of kaon $\beta$
decays, according to (\ref{K02pi}), comes from the experimental
accuracy of the branching ratio (\ref{Kp2pi}) of hadronic mode of
the charged kaon decays $\delta B/B\approx 6\times 10^{-3}$. Its
contribution into $V_{ud}$ error, $3\times 10^{-3}$, is also
competitive with pion $\beta$ decay uncertainty. It may be
improved in future, because the last direct measurement of this
ratio has been done more than thirty years ago~\cite{Chiang}.

Speaking about the experimental selection of the rare process
(\ref{Kdecay}) we should note that the main background to it comes
from $K_{e4}$ decay with branching ratio $(5.21\pm 0.11)\times
10^{-5}$~\cite{NA48}. The difference in  $\Delta S = \Delta Q$
selection rule can be used for discrimination of these decays
in the case of tagged  $K^0$, $\bar{K}^0$ beams. However, it
cannot help in the case of $K_L$ beam, containing both $K^0$
and $\bar{K}^0$ meson states. Nevertheless, these decays can
be well separated kinematically. First of all, the two
pions in the final state of (\ref{Kdecay}) come with definite
invariant mass $M_{\pi\pi}=m_{K^+}$ from the two-particle $K^+$ decay
and apart from $K_L$ decay point. In the same time $K_{e4}$ process is
three body decay and its distribution in $M_{\pi\pi}$ variable has
a continuous spectrum with the maximum at 340 MeV and follows
decreasing to the end. Good knowledge of the form
factors~\cite{Pais} of $K_{e4}$ process allows us to subtract
background under the peak from process (\ref{Kdecay}) around
$m_{K^+}$.

The last and the main source of the uncertainty in $V_{ud}$
determination comes from the $K^0-K^+$ mass difference (\ref{DK}),
which enters into the rate of kaon $\beta$ decay (\ref{tKbeta}) in
the fifth power. It leads to inappropriate contribution to $\delta
V_{ud}/V_{ud}\approx 1.7\times 10^{-2}$. Experimental situation at
present resembles the one in the 1986 for $m_{\pi^+}-m_{\pi^0}$
pion mass difference, when its uncertainty was almost completely
determined by the uncertainty in the neutral-pion mass
measurements~\cite{Dpi}. Taking into account the recent
considerable progress in kaon study, one hopes that new
measurements of $K^0$ mass~\cite{K0mass} and directly $K^0-K^+$
mass difference will be made.

We did not discuss the radiative corrections $\delta$ and their
uncertainties for kaon $\beta$ decay. Most probably they can be
calculated in the same lines as for pion $\beta$
decay~\cite{Cirigliano}.

In this letter we have proposed the theoretical possibility to
extract $V_{ud}$ matrix element from kaon $\beta$ decay. We have
given only primeval experimental insights for the registration of
this process and the estimation of the main uncertainties. Of
course, in order to provide an experimental realization of this
project, a systematical study of such a project is necessary. It
will be interesting to analyze the possibility of such
measurements within the proposed project~\cite{Hsiung} searching
$K_L\to\pi^0\nu\bar{\nu}$ decay.

\section*{Acknowledgements}

I would like to thank C. Cheshkov, V.~Genchev and L.~Widhalm
for the useful comments.
I am grateful to D. Kirilova for the overall help
and appreciate the stimulating ICTP environment.

\pagebreak[3]

\end{document}